\documentclass[11pt]{article}

\usepackage{booktabs} % For formal tables
\usepackage{natbib} % for bibliography
\usepackage{luca} % My defs
\usepackage{graphicx}
\usepackage{amsfonts,amsmath,amssymb}
\usepackage{url}
\usepackage{hyperref}
\usepackage{comment} % To comment out blocks of text
\usepackage{float} % To put figures/tables exactly where I want them
\usepackage{tablefootnote} % To add footnotes below tables
\usepackage{xcolor}
\usepackage{geometry} 
\usepackage{lipsum}

\makeatletter
\renewcommand{\maketitle}{\bgroup\setlength{\parindent}{0pt}
\begin{flushleft}
  \textbf{\Large\@title}
  \vskip 5mm
  \@author
  \vskip 1mm  
  \textbf{(\url{https://truthvalue.org})}
\end{flushleft}
\vskip 1mm
\begin{center}\@date\end{center}
\vskip 5mm
\egroup
}
\makeatother

\def\mybar#1{\kern-3mm\color{red}\rule{#1cm}{8pt}\kern-#1cm\kern+3mm\color{black}}

\begin{document}
\title{Reputation Systems for News on Twitter: {A} Large-Scale Study}

%\author{Luca de Alfaro}
%\authornote{Dr.~Trovato insisted his name be first.}
%\affiliation{%
%  \institution{UC Santa Cruz}
  % \streetaddress{P.O. Box 1212}
%  \state{California, USA}
%  % \postcode{43017-6221}
%}
%\email{luca@ucsc.edu}

\author{Luca~de~Alfaro$^1$, Massimo~Di~Pierro$^2$, Rakshit~Agrawal$^3$, Eugenio~Tacchini$^4$, Gabriele~Ballarin$^5$, Marco~L.~Della~Vedova$^6$, Stefano~Moret$^7$ \\
{\footnotesize \vskip 2mm
1. Department of Computer Science, UC Santa Cruz, CA, USA. {\tt luca@ucsc.edu} \\
2. School of Computing, DePaul University, Chicago, IL, USA. {\tt massimo.dipierro@gmail.com} \\
3. Department of Computer Science, UC Santa Cruz, CA, USA. {\tt ragrawa1@ucsc.edu} \\
4. Universit\`a Cattolica, Piacenza, Italy. {\tt eugenio.tacchini@unicatt.it} \\
5. Independent researcher. {\tt gabriele.ballarin@gmail.com}\\
6. Universit\`a Cattolica, Brescia, Italy. {\tt marco.dellavedova@unicatt.it} \\
7. \'Ecole Polytechnique F\'ed\'erale de Lausanne, Switzerland. {\tt moret.stefano@gmail.com}
}}

% Please, add your name here, oh other authors. 

%\renewcommand{\shortauthors}{}

\date{Jan 8, 2018}

\maketitle

\begin{abstract}
Social networks offer a ready channel for fake and misleading news to spread and exert influence.
This paper examines the performance of different reputation algorithms when applied to 
a large and statistically significant portion of the news that are spread via Twitter. 
Our main result is that simple algorithms based on the identity of the users spreading the news, as well as the words appearing in the titles and descriptions of the linked articles, are able to identify a large portion of fake or misleading news, while incurring only very low ($<1\%$) false positive rates for mainstream websites.
We believe that these algorithms can be used as the basis of practical, large-scale systems for indicating to consumers which news sites deserve careful scrutiny and skepticism.
\end{abstract}

% \keywords{fake news, hoax detection, social networks, machine learning}

% Content
\section{Introduction}

Social networks such as Facebook and Twitter offer fake news the means to spread, and the resonance chambers where users can consume and reshare them \cite{allcott_social_2017}. 
% The reach that fake news achieve in turn encourages the creation of ever more fake news.
This situation has prompted the research community to create the means to identify and label fake news as they spread across social networks; \cite{shu_fake_2017} provides an excellent and up to date overview of such research. 

Previous experiments with fake news detection focused on curated sets that captured only a portion of what is shared.
Our interest here lies in the question of how fake news detection systems perform when applied to the full variety of news being shared via Twitter.
What performance can be obtained?
How do the algorithms behave on news from well-known websites? 
How effective are the methods in unearthing previously unknown websites that spread fake or misleading news? 

To answer these questions, over a period of several months, we collected a large subset of URLs shared on Twitter pointing to news articles, including all the news posted on mainstream sites and agencies, all the news from selected scientific and peer reviewed publications, all the news from hundreds of sites that have appeared in published curated lists of sites with dubious reputation or low editorial controls, and all news being spread by a selected group of Twitter users we chose to follow. 
This initial set was then augmented with the result of queries aimed at dataset completion, finding additional users who shared the news, and additional news shared by users in our dataset.  
The result is a dataset consisting of 5.5 million news articles, 88 million tweets, and 9 million users.
% This constitutes what we believe to be a statistically significant portion of the news spread via Twitter from August to November 2017.
For each news article, we collected both the item title and description, and the identities of the users who shared the item. 
Our experiments are carried out over temporal slices generally consisting of 1.4 million news and 20 million tweets.

We experimented with three algorithms for identifying fake or misleading news. 
The first algorithm, from \cite{tacchini_like_2017}, uses the users that spread a piece of news, and optionally the title and description words, as features, and builds a logistic-regression classifier. 
The second algorithm is a variant in which the user identities are first aggregated via topic modeling \cite{blei_latent_2003}. 
The third algorithm is a graph-based crowdsourcing algorithm from \cite{de_alfaro_reliable_2015,tacchini_like_2017}.
All of these algorithms are well suited for large-scale implementation.
Our ground truth consists in two independently developed lists of low quality news sites: the Opensources list \cite{opensources_curated_2017}, and a list from Metacert\footnote{http://www.metacert.com}; each list comprises about 500 sites. 

Our main result consists in showing that the algorithms can be adapted, trained, and tuned to successfully discover a significant portion of the low-quality news in our test dataset (above 50\% by default, or above 85\% when limiting the comparison to fake news with more than 10 tweets, which constitute the most ``viral'' news), while incurring a false positive rate on mainstream news below 1\%. 
In preliminary user trials, we observed this low false positive rate to be essential: mainstream reliable media such as the major newspapers produces a very large amount of news daily, and even a small fraction labeled erroneously as fake translates in a large number of incorrectly flagged news, eroding user confidence in the system. 
% This detection performance is achieved by training the logistic-regression algorithms on news that were shared more than once: disregarding once-only shared news reduces somewhat the detection ability, but is instrumental in achieving the low false positive rate.
We provide extensive results on how the algorithms perform on individual news sites, both mainstream and lesser known.

Furthermore, we show that these algorithms are able to generalize from known low-quality news sites to other sites that deserve to be scrutinized. 
In particular, we show that starting from one of our ground truths, we can recover a majority of the sites belonging to the other ground truth.

Together, these results indicate that the algorithms can perform well when applied to the full set of news items shared via Twitter, an essential milestone for their wider adoption.
We like to think of these algorithms as {\em reputation systems} for news, rather than proper fake news detection algorithms.
These algorithms are too shallow semantically to be able to actually discern genuine from fake news; this is a task best carried out by humans, or by systems performing a much deeper analysis of the text, and requires taking into account a broader context.
Our aim is more modestly to show that reputation systems with high predictive value and low false positive rate on mainstream news are achievable in practice.

The system for fake news detection described in this paper has been implemented as part of the Truth Value Project, and it is available as a service at the URL \url{https://truthvalue.org}
The Truth Value Project system continuously collects information from Twitter, and it allows uses to query the system for the reputation of any arbitrary news article, identified by its URLs. 
Users can also vote news articles as being reliable, or fake/misleading; the users' opinions are weighted (positively or negatively) depending on their reputation. 
Additionally, we offer two other ways to interact with our system: we provide a Twitter bot ({\tt hoaxbot}) that can respond to enquiries from the users, and a browser bookmarklet which enables users to quickly access the scores of sites they are accessign with their browser.

\section{Related Work}

Fake news is a phenomenon that has received wide attention of late, and motivated much research aimed at detecting fake news automatically; a comprehensive and recent survey is given in \cite{shu_fake_2017}. 
Many approaches are based on the content of the news.
Some of these approaches find their roots in the much older work aimed at detecting spam; see for instance \cite{mason_filtering_2002,vukovic_intelligent_2009}. 
\cite{wang_liar_2017} uses deep learning, LSTMs, and SVMs to classify about 12,000 sort text statements from Politifact; \cite{riedel_simple_2017,ahmed_detection_2017} used a text TF-IDF model followed by a neural net to classify a few thousands text items. 
The writing style has been used in \cite{potthast_stylometric_2017} to classify 1,600 items from BuzzFeed. 

An extensive study was performed in \cite{castillo_information_2011}, which collected tweets verging on 2,500 topics that were trending on Twitter between April and July 2010.  
The topics were classified on the basis of the features of the users and the tweets belonging to the topics, showing high classification accuracy for the topics; the best-performing classification algorithm was based on a decision-tree classifier. 
The social dynamics on Twitter and the temporal propagation of posts has been the subject of a very extensive study in \cite{shao_spread_2017}, with the aim of quantifying the effect of bot accounts, detecting them, and tracing how the information spreads. 
The study is similar to ours in extent, involving millions of news items and Twitter users.

Temporal aspects of news spreading have been studied and correlated to the reliability of news in \cite{ma_detect_2015}, where about a million tweets are used to analyze a hundred news items, and in \cite{shao_hoaxy:_2016}, where large-scale measurements are provided. 

The present work is inspired by work on reputation systems, in which the specific identity of users is used as a feature. 
Some notable recent work in this direction is \cite{tacchini_like_2017}, some of whose algorithms we study in this paper, and \cite{ruchansky_csi:_2017}, where user identities are fed along with text into a deep-learning system.

\section{Dataset}

% Our goal is to study and compare the performance of reputation systems for news spread via Twitter. 
% We identify news with URLs, and the reputation systems we consider aim at associating a value of reputation to URLs spread via Twitter.

\subsection{Dataset composition}

To characterize the performance of reputation systems for news spread via Twitter, it is essential to have a dataset that covers the full spectrum of news that are shared on Twitter, from mainstream sites, to sites whose news quality is less certain.  
Since our concern is news reputation, rather than individual tweet reputation, we identify news with URLs, and we restricted our attention to tweets that contain at least one URL. 

Ideally, we would like our dataset to include all URL-containing tweets and construct the full bipartite graph with URLs on one side, and Twitter users mentioning those URLs on the other side. 
However, acquiring this much data would be prohibitively expensive.
We therefore settled on a sampling strategy that allowed us to collect a large number of news-related tweets and re-tweets, and build a significative sub-graph.
We achieved this by gathering URL-containing tweets using the Twitter streaming API. We collected all Tweets from August 1, 2017 until December 8, 2017 that satisfied at least one of the following filters:
\begin{itemize}

\item {\em Mainstream news:} all tweets containing URLs from a list of mainstream news sites, including ABC News, Breitbart, BuzzFeed, CBS News, Channel 7 News, CNN, Fox News,  MSNBC, NBC News, The Huffington Post, The Economist, The Guardian, The Hill, The Onion, The New York Post, The New York Times, The Times, The US Herald, The Washington Post, USA Today, and US News;
 
\item {\em News agencies:} all tweets from The Associated Press and Reuters;
 
\item {\em Scientific, and peer-reviewed news:} all tweets from ArXiV, Nature, and Science Magazine;

\item {\em Fact-checking sites:} all tweets from Politifact and Snopes;
 
\item {\em News from selected users:} all tweets from a set of about 160 users, which included many of the most active tweeters in news and science; and 
 
\item {\em Low-quality news:} all tweets that mentioned URLs from many of the sites mentioned as fake, bias, unreliable, clickbait, or consipiracy in the lists \cite{ZimdarsFalseMisleadingClickbaity2016,opensources_curated_2017}.  The original lists consisted of 1,000 sites; we excluded many that seemed to be no longer active, or that published news only sporadically.

\end{itemize}
In addition, we ran specific queries aimed at completing the graph, by periodically querying (within the limits of the Twitter API) for URLs that were mentioned only once or twice, in the attempt to find more tweets referencing these URLs; and users from whom we had a limited number of tweets, in the attempt to discover further activity from such users. 

Overall, between the streaming and the graph-completion queries, we gathered about a million URL-containing tweets per day. 
Upon gathering each tweet, we parsed it, and we downloaded the HTML of the reference page.
From the semantic tags ({\tt og:url}, {\tt og:title}, and {\tt og:description}) in the HTML we extracted and stored the canonical URL, the title, and the description. 

We report in Table~\ref{table-dataset-composition} the list of news sites with the most URLs in the dataset for the period from September 1 to November 30, 2017.
% In Figure~\ref{fig-testing-histogram} we give the histogram  of the number of shares (tweets) per URL in the period we use to measure the algorithm performance, running from October 31 to November 26, 2017. 
% As expected, most URLs have only one share; however many URLs of news that went viral had well over 10,000 shares. 

\begin{table}
\begin{center}
\begin{tabular}{l|r||l|r}
Site & \% & Site & \% \\ \hline
                   youtube.com  & 4.319 &                  wordpress.com  & 0.754  \\
                   nytimes.com  & 3.145 &                     nypost.com  & 0.625  \\
               theguardian.com  & 2.904 &                    thehill.com  & 0.619  \\
            huffingtonpost.com  & 1.964 &                    latimes.com  & 0.616  \\
            washingtonpost.com  & 1.944 &                  breitbart.com  & 0.609  \\
                     arxiv.org  & 1.585 &                    cbsnews.com  & 0.563  \\
                  usatoday.com  & 1.504 &                    reuters.com  & 0.426  \\
                indiatimes.com  & 1.458 &                     reddit.com  & 0.388  \\
                   foxnews.com  & 1.262 &                dailycaller.com  & 0.367  \\
                  blogspot.com  & 1.202 &                    newsmax.com  & 0.336  \\
\end{tabular}
\end{center}
\caption{The 20 news sites with the most URLs in our dataset in the period from September 1 to November 30, 2017. }
\label{table-dataset-composition}
\end{table}

% \begin{figure}
% \centering
% \includegraphics[scale=0.7]{figs/reduced_test_degree.pdf}
% \caption{Histogram of number of tweets per URL in our testing dataset; the histogram is truncated at 30 tweets. 
% The histogram is obtained from the subsample of tweets that appeared from October 31 to November 26 on alternating days.}
% \label{fig-testing-histogram}
% \end{figure}

Once the tweets and the URLs had been obtained, we processed the dataset and we associated with each URL the following information:
\begin{itemize}
\item {\em URL date:\/} the date at which we first saw the URL. 
\item {\em Users:\/} the list of Twitter usernames who shared the URL. 
\item {\em Title:\/} the title of the news article, as given in the {\tt og:title} social tag.
\item {\em Description:\/} the description of the news article, as given in the {\tt og:description} tag.
\end{itemize}

\subsection{Ground Truth}

All the machine learning algorithms that we utilize follow into the category of semi-supervised learning algorithms. In order to train them we need a set of example URLs that we can reliably categorize as good (reliable news) or bad (fake and misleading news). We refer to this set of categorized URLs as {\em ground truth}.
 
Ideally we would like to utilize a large, unbiased, and possibly crowdsourced, ground truth.
Unfortunately, such a set is not currently available to the authors. 
Most sets of crowdsourced URLs are fairly small, generally numbering less than 1,000 URLs.
It is difficult to build, on the basis of such small sets, algorithms that generalize well to the whole set of news being shared on Twitter.
Furthermore, and perhaps even more importantly, if the testing set is small it does not yield useful information on the algorithms' performance because we cannot extrapolate the results to the full set of URLs collected on Twitter, which is the question that interests us here.
 
As the next-best solution, we use as ground truth, two curated lists of news sites publishing (in the opinions of the curators) fake, misleading, unchecked, and biased news: the {\em Metacert\/} list, and the {\em Opensources\/} list.
\begin{itemize}

\item{\em Opensources list.}
The {\em Opensources\/} list consists of sites derived from a list originally posted by Melissa Zimdars \cite{opensources_curated_2017}.
We kept all sites in that list with type {\em bias, fake, junksci, hate, conspiracy,} or {\em clickbait.}
This resulted in 581 sites.

\item{\em Metacert list.}
The {\em Metacert\/} list consists of sites flagged as fake, misleading, or unreliable news sources using a crowdsourcing method by Metacert\footnote{https://metacert.com/}, and kindly shared with the authors. 
The list consists of 500 sites.

\end{itemize}
These two lists have partial overlap, sharing 331 sites. 

Using these two lists as ground truth entails two problems. 
First, the lists consist of sites, not of URLs. 
There is thus the risk that our algorithms simply learn the sites, rather than the characteristics of fake news. 
We have taken two steps to counter this. 
As mentioned above, we ensure that no information used in the machine learning directly mentions the site. 
In particular, we remove all mentions of the sites where the news appear (and their aliases) from the content we use as ML feature. 
Further, we will present extensive results on {\em cross-learning,} in which we use one set of sites (for instance, the Metacert list) to infer the reliability of URLs belonging to another set of sites (for instance, the Opensources list).

The second problem is that the Metacert and Opensources lists are only partial lists of sites containing unreliable information.
Even when we use the union of the two lists of sites, we certainly cannot assume that all URLs not belonging to these sites contain reliable news. 
In other words, the {\em false positives\/} of our machine learning --- the URLs that are labeled as low reputation, but do not belong to a site in the list --- may not be mistakes, in the sense that they may consist in part of URLs that, upon examination, reveal themselves to be fake or misleading news. 
For this reason, when reporting the performance of the ML algorithms, we will not blindly accept the classification given by the ground truth. 
Rather, we will report detailed results on which news sites are reported to contain low-reputation news, and in what proportion. 
As we will see, the ML algorithms we study in general have the useful ability to discover fake and unreliable news beyond those that belong to the ground truth.

\begin{figure}
\begin{center}
\includegraphics[scale=0.32]{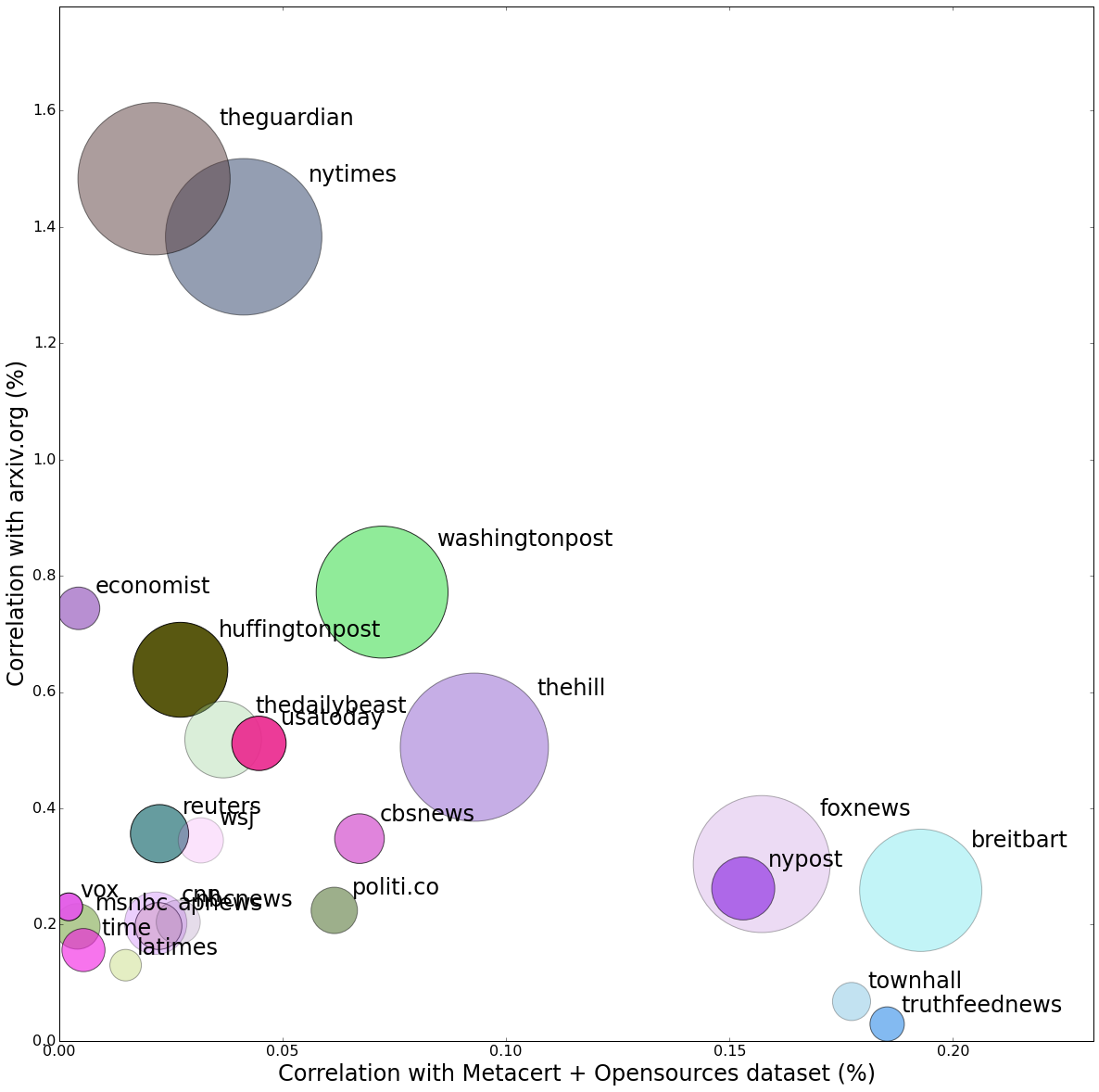}
\end{center}
\caption{Average correlation of most tweeted news sites with known fake news sites (Opensources + Metacert lists) vs correlation with science related (arxiv.org).  The area of each circle is proportional to the number of URLs we have from each site. The data is from November 1 to November 7, 2017.}
\label{fig-circles}
\end{figure}

To give an idea of the breadth of news collected, we give in Figure~\ref{fig-circles} the correlation between the main news sites, and the sites appearing in the union of the Opensources and Metacert lists on one side, and the scientific site arxiv.org on the other. 
We compute the correlation between two sites $i,j$ as $T_{ij}/\sqrt{T_i T_j}$ where $T_{ij}$ counts the number of users that tweeted about both domains weighted by their number of tweets; $T_i$ and $T_j$ count the number of users that twitted about domain $i$ and $j$ respectively, also weighted by the number of tweets. 

\section{Methods}

We build reputation systems for news shared on Twitter following three different approaches. 
In the first approach, we consider each user who shared the news article, and each word that appears in the title and description of the article itself, as an individual feature.  
This gives rise to a large number of features, on which we train a simple logistic regression model \cite{tacchini_like_2017}. 
In the second approach, we aggregate users and title/description words using topic analysis, and we then train neural network classifiers based on the resulting topics. 
In the third approach, we implement the ``harmonic'' crowdsourcing algorithm described in \cite{de_alfaro_reliable_2015,tacchini_like_2017}.

\subsection{Logistic-regression classifier based on users and words}
\label{sec-logistic}

\subsubsection{Architecture} 

Our first and simplest model is from \cite{tacchini_like_2017}: we use as list of features for a piece of news (a URL after canonicalization) the individual users who shared the URL by tweeting or retweeting it, alongside with the words that appear in the title and description of the news article itself. 
This gives rise to a large set of features, numbering in the hundred of thousands: each user and each word is a feature. 

On this large, and rather sparse feature set, we train a simple logistic regression model. 
Logistic regression models are well suited to coping with large and sparse feature sets. 
Intuitively, the coefficient associated with each user is positive if the user tends to share high-reputation news, and negative if the user tends to share low-reputation news. 
Similarly, the coefficient of a word is positive if the word tends to appear in the title or description of high-reputation news, and negative if the word tends to appear in the title or description of low-reputation news. 
% Thus, the logistic-regression classifier provides an extremely simple way of attaching positive and negative influences to users and words, and obtaining the reputation of a news URL by adding the influences.

We consider three kind of models: 
\begin{itemize}
\item {\bf LR-U}, built on users only.
\item {\bf LR-UT}, built on users and title/description text.
\item {\bf LR-T}, built on title/description text only.
\end{itemize}
The models are all implemented using the {\tt scikit-learn\/} Python package. 
Since our ground truth is based on sites, rather than individual URLs, we removed from the title and descriptions all mentions of the news site to which the individual URLs belong, to avoid having a ``cheat factor'' in the experiments.

\subsubsection{Training and testing datasets}

To evaluate the logistic-regression based classification methods, we split the dataset into {\em training\/} and {\em testing\/} datasets, as follows.
\begin{itemize}
\item The full training dataset consists of all URLs that have been first seen from September 1st, 2017, to October 20, 2017. 
For each URL, we keep all the tweets whose timestamp is strictly before October 30, 2017.

\item The full testing dataset consists of all URLs that have been first seen from October 31, 2017, to November 26, 2017. 
For each URL, we keep all the tweets that mention it.
\end{itemize}
The construction of the full training and testing datasets ensures that the URLs they contain are disjoint, and that the testing dataset is built only from information that chronologically precedes all of the information in the testing dataset. 
The interval from October 20 to October 30, 2017, allows us to collect tweets mentioning the URLs seen first on October 20 for 9 additional days, until midnight of October 29 (all dates used in the experiment are in UTC). 

As the full training and testing datasets are quite large, we subsampled them by keeping only the tweets appearing on alternating days, starting from September 1, and October 31, respectively. 
Furthermore, we obtained a {\em min-2 training dataset\/} by considering only URLs that were retweeted at least twice; this weeded out from the training set many low-significance URLs, and let to higher classification precision for mainstream news, as we will see. 
The characteristics of the resulting datasets are reported in Table~\ref{table-reduced-datasets}.

\begin{table} 
\begin{center}
\begin{tabular}{l|r|r}
Dataset & URLs & Tweets \\ \hline
Training dataset & 787,601 & 14,587,984 \\
Min-2 training dataset &  275,400 & 14,075,783 \\
Testing dataset & 607,299 & 7,967,170
\end{tabular}
\end{center}
\caption{Sizes of training and testing datasets for logistic-regression based classifiers.}
\label{table-reduced-datasets}
\end{table}

The composition of the two ground truths is summarized in Table~\ref{table-gt-sizes}.
As the number of positive and negative instances in the training set is highly unbalanced (in the training dataset, only 21,263 of 787,601 URLs are considered fake news), we train the logistic classifier giving class weights that are inversely proportional to class sizes, thus minimizing per-class errors. 

\begin{table}
\begin{center}
\begin{tabular}{l|r|r|r|r}
& \multicolumn{4}{|c}{Number of URLs} \\
& Opensources & Metacert & Common & Total \\ \hline
Train & 21,263 & 21,146 & 14,339 & 787,601 \\
Min-2 train & 13,213 & 13,305 & 9,104 & 275,400 \\
Test  & 13,388 & 11,740 & 9,210 & 607,299 \\
\end{tabular}
\end{center}
\caption{Number of URLs in the Opensources and Metacert  lists that appear in our training and testing sets for logistic-regression classifiers, alongside the total number of URLs in the datasets.}
\label{table-gt-sizes}
\end{table}

\subsection{Topic-based models}

A limitation of the above logistic-based classification model is that we can assign coefficients only to users and words that we have seen in our ground truth. 
In particular, if a user $u'$ has been recently sharing similar URLs as users $u_1$ and $u_2$, and we have extensive data on the quality of the URLs shared by $u_1$ and $u_2$ but not by $u'$, we still cannot attribute a coefficient (a positive or negative influence) to $u'$.
Topic-based models extract the common behavioral similarities among users, summarizing them in a number of {\em topics\/} that each user is likely to share. 
We can then train our models on the basis of topics, rather than individual users. 
In the example above, the topic preference of $u'$ would be close to the one of $u_1$ and $u_2$, so if we knew how the topic preference of $u_1$ and $u_2$ and how those topics are associated to high or low reputation news, we could transfer this information to $u'$. 

\subsubsection{Architecture}
\label{sec-training-topic-modeling}

We create our topic models by following the classical Latent-Dirichelet Allocation models (LDA) for text \cite{blei_latent_2003}.
We considered each URL as a document, with the users that shared that URL as the words, and we applied standard LDA-based topic-modeling to the resulting document corpus via Gensim \cite{rehurek_lrec}. 
We constructed the topic modeling on the joint dataset consisting of {\em both\/} training and testing data.  
Constructing topic models is a fully unsupervised operation which can be easily repeated periodically (e.g., once a day), so that in practice it is feasible to maintain an updated topic model encompassing the whole dataset. 
Constructing the topic modeling on the joint dataset enabled us to exploit behavior similarity between the users in the training dataset, and those appearing only in the testing dataset.

Once the topic model was constructed, we used Gensim to map each URL, with its sharing users, to a topic vector.
We first experimented with training a logistic classifier on topic vectors, but this worked rather poorly. 
The topics provide rich semantical information, and apparently it is necessary to consider the conjunctions, and disjunctions, of topics in order to arrive at a good classification of news URLs. 
Thus, after some experimentation, we settled on a neural-net classifier with one 100-neuron hidden layer with rectified liner activation, and one final neuron outputting the classification.
Again, due to class imbalance between the true and fake classes, in the learning we use class weights that are inversely proportional to class size. 

\subsubsection{Training and testing datasets}

Due to computational limitations (topic decomposition is an expensive operation), we used somewhat smaller datasets than the ones previously described.
Precisely:
\begin{itemize}
\item The raw training datasets consists of all URLs with at least two tweets that were first seen in 2017 on September 1, 5, 9, 13, 17, 21, 25, 29, and October 3, 7, 11, 15, 19.  This dataset contained 144,137 URLs, and encompassed 7,395,756 tweets. 
\item The raw testing dataset consisted, as in Section~\ref{sec-logistic}, in all URLs first seen on alternating days, starting on October 31, and ending in November 26, 2017. This dataset contained 607,299 URLs, encompassing 7,967,170 tweets. 
\end{itemize}
To further simplify the problem computationally, we noted that URLs that are shared only a few times, and users who share only occasionally, do not contribute much information to the topic factorization. 
Hence, we computed the topic factorization on the basis of URLs and users in the union of the raw training and testing datasets that appeared at least 5 times.

Once the topic modeling was computed, we used our raw training set above to train the neural net on the basis of the URL topics.  As testing set, again to simplify the problem computationally ,we considered a random subsampling of 20\% of the raw testing set mentioned above, resulting in 121,460 URLs. 
The compositions of the training and testing sets is given in Table~\ref{table-gt-sizes-topics}.

\begin{table}
\begin{center}
\begin{tabular}{l|r|r|r|r}
& \multicolumn{4}{|c}{Number of URLs} \\
& Opensources & Metacert & Common & Total \\ \hline
Train & 7,069 & 7,032 & 4,876 & 144,137 \\
Test  & 2,664 & 2,331 & 1,810 & 121,460 \\
\end{tabular}
\end{center}
\caption{Number of URLs in the Opensources and Metacert  lists that appear in our training and testing sets for topic modeling. 
The different proportion of URLs that belong to the Opensources and Metacert sets, compared to the total URLs, depends on the fact that the training set consists only of URLs that were shared at least twice.}
\label{table-gt-sizes-topics}
\end{table}

\subsection{Crowdsourcing-based models}

The third method we applied is based on the crowdsourcing-inspired ``harmonic boolean label crowdsourcing'' model of \cite{de_alfaro_reliable_2015,tacchini_like_2017}. 
As for topic modeling, the crowdsourcing models are developed on the entire dataset; indeed, the system we use for Twitter data acquisition is able to run the harmonic model computation every few hours, ensuring that up-to-date results are always available. 

In boolean label crowdsourcing models, users provide True/False labels for a set of items, and the crowdsourcing algorithms then attribute to each item an estimate of truth or falsehood, and to each user an estimate of how much the user is likely to tell the truth or to lie \cite{karger_iterative_2011,liu_variational_2012}.
As in \cite{tacchini_like_2017}, we use a variant of these methods in which the truth or falsehood of a subset of items is taken as known; the algorithms then propagate the information to the remaining users and items. 
We use the {\em harmonic} algorithm of \cite{de_alfaro_reliable_2015}, both because it is efficient on very large-scale data, and because it has been shown to give good results in \cite{tacchini_like_2017}.

\subsubsection{The harmonic algorithm}

We represent our entire dataset as a bipartite graph $(I \union U, T)$, where $I$ are the news items, $U$ are the users, and $T \subs I \times U$ are the shares, so that a share $(i,u)$ represents user $u$ tweeting item $i$.
We denote by $\partial i = \set{u \mid (i, u) \in L}$ and $\partial u = \set{i \mid (i, u) \in L}$ the 1-neighborhoods of an item $i \in I$ and user $u \in U$, respectively.

The harmonic algorithm maintains for each node $v \in I \union U$ two non-negative parameters $\alpha_v$, $\beta_v$ defining a beta distribution: intuitively, for a user $u$, $\alpha_u - 1$ represents the number of times we have seen the user share a reliable news item, and $\beta_u - 1$ represents the number of times we have seen the user share an unreliable item.
Similarly, for a news item $i$, $\alpha_i - 1$ is the number of shares from reliable users, and $\beta_i - 1$ the number of shares from unreliable users. 
For each node $v$, let $q_v = (\alpha_v - \beta_v) / (\alpha_v + \beta_v)$: for a news item $i$, positive values of $q_i$ indicate a likely reliable news, and negative values, a likely hoax. 

Let the training set consist of two subsets $I_F, I_N \subs I$ of fake and non-fake news. 
The algorithm sets $q_i := -1$  for all $i \in I_F$, and $q_i := 1$ for all $i \in I_N$; it sets $q_i = 0$ for all other posts $i \in I \setm (I_F \union I_N)$.
The algorithm then proceeds by iteratively propagating the information from items to users, and from users to items. 
First, for each user $u \in U$, it lets: 
\begin{align*} 
    \alpha_u & := c + \sum \set{q_i \mid i \in \partial u, q_i > 0} \\
    \beta_u & :=  c - \sum \set{q_i \mid i \in \partial u, q_i < 0} \eqpun . % \\
%    q_u & := (\alpha_u - \beta_u) / (\alpha_u + \beta_u) \eqpun .
\end{align*}
The positive constant $c = 0.02$ acts as regularization. 
In the second part of the iteration, the algorithm updates the values for each item $i \in I \setm (I_F \union I_N)$ (without affecting the ground truth) by:
\begin{align*} 
    \alpha_i & := c + \sum \set{q_u \mid u \in \partial i, q_u > 0} \\
    \beta_i & :=  c - \sum \set{q_u \mid u \in \partial i, q_u < 0} \eqpun . % \\
%    q_i & := (\alpha_i - \beta_i) / (\alpha_i + \beta_i) \eqpun .
\end{align*}
We experimented with various numbers of iterations, and we settled on using four iterations, which seem sufficient to spread information from the labeled nodes.
Once the iterations are concluded, we classify a news item $i$ as fake  if $q_i < 0$, and as reliable otherwise; thus, we take $q_i$ as the reputation of $i$.

\subsubsection{Training and testing datasets}

The harmonic algorithm, as mentioned, is run on our entire graph of URL and tweets; the graph consists to date of 5.5 million URLs and 88 million tweets.
We use as training set the approximately 2.5 million URLs collected before October 15, 2017, and as testing set, the approximately 2.5 million URLs collected after November 1st, 2017.
In the training set, we take as negative ground truth all URLs that belong to the Opensources or Metacert ground truths; let $n$ be the number of such URLs. 
As $n \ll 2,500,000$, if we labeled positively all the remaining URLs in the training set, the training set would have a strong class imbalance.
The harmonic algorithm does not come with a ``control knob'' comparable to class weight normalization for logistic regression.
To obtain a balance between the number of URLs labeled positively and negatively for training, we sample a subset of URLs not in the Opensources and Metacert lists, and we will label only those positively. 
We consider two subsamplings: the {\em 1x subsampling,} in which we subsample $n$ URLs, and the {\em 4x subsampling,} in which we subsample $4n$ URLs.

\section{Results}

\begin{table*}
\begin{center}
\tiny\begin{tabular}{lr||r|r|r||r|r|r||r||r|r}
& & \multicolumn{3}{c||}{Full Training} & 
   \multicolumn{3}{c||}{Min-2 Training} & & \multicolumn{2}{c}{Harmonic} \\
& & LR-U & LR-UT & LR-T & LR-U & LR-UT & LR-T & Topics & \multicolumn{1}{c|}{1x} & \multicolumn{1}{c}{4x} \\ \hline
{\em All URLs}
& Hoax recall:    & 57.64 & 61.14 & 57.25 & 46.84 & 53.25 & 53.21 & 54.80 & 91.20 & 74.34 \\
& Nonhoax recall: & 97.40 & 97.75 & 94.18 & 98.76 & 98.51 & 94.82 & 93.88 & 89.74 & 96.69 \\
& Hoax precision: & 33.30 & 38.15 & 18.15 & 46.14 & 44.61 & 18.81 & 16.50 & 16.64 & 32.87 \\ \hline
{\em URLs with $\geq$2 shares}
& Hoax recall:    & 72.12 & 71.69 & 62.69 & 63.94 & 66.76 & 59.19 & 66.56 & 91.20 & 74.34 \\
& Nonhoax recall: & 95.62 & 96.66 & 91.73 & 97.55 & 97.84 & 93.25 & 92.23 & 89.74 & 96.60 \\
& Hoax precision: & 41.44 & 49.19 & 24.54 & 46.14 & 57.07 & 27.34 & 26.60 & 16.64 & 32.87 \\ \hline
{\em URLs with $\geq$5 shares}
& Hoax recall:    & 82.57 & 81.27 & 67.38 & 79.03 & 78.94 & 64.93 & 76.32 & 91.83 & 69.79 \\
& Nonhoax recall: & 95.47 & 96.66 & 90.14 & 97.00 & 97.49 & 92.24 & 92.83 & 85.44 & 95.32 \\
& Hoax precision: & 50.24 & 57.43 & 27.46 & 52.83 & 63.56 & 31.66 & 36.12 & 24.79 & 43.75 \\ \hline
{\em URLs with $\geq$10 shares}
& Hoax recall:    & 87.14 & 85.43 & 69.90 & 85.43 & 84.80 & 68.06 & 83.13 & 93.52 & 70.83 \\
& Nonhoax recall: & 95.64 & 96.63 & 88.91 & 96.77 & 97.33 & 91.50 & 93.20 & 84.54 & 95.43 \\
& Hoax precision: & 54.28 & 60.16 & 27.24 & 61.10 & 65.37 & 32.23 & 40.70 & 25.44 & 46.62 \\ 
\end{tabular}
\end{center}

\caption{Hoax and non-hoax recall, and hoax precision, expressed as percentages, for the news reputation systems compared in this paper. 
The methods are LR-U: logistic regression based on users; LR-UT: logistic regression based on users and text; LR-T: logistic regression based on text; Topics: topic analysis, and Harmonic: harmonic crowdsourcing algorithm.
For logistic regression, we report results for two training sets: the full one, and the one consisting of URLs that have been shared at least twice.
For the other methods, we only report results on the full training set.
For Harmonic, we report the results with both 1x and 4x sampling of good URLs. 
The results are based on the Opensources ground truth.
}
\label{table-res-summary}
\end{table*}

Table~\ref{table-res-summary} summarizes the results for the various algorithms under consideration. 
For each algorithm, and each test-set, we report: 
\begin{itemize}

\item {\bf Hoax recall:} the percentage of hoax news in the testing dataset that the algorithm correctly labels as hoaxes.

\item {\bf Nonhoax recall:} the percentage of non-hoax news in the testing dataset that the algorithm correctly labels as non-hoaxes.

\item {\bf Hoax precision:} the percentage of news in the hoax-labeled news in the testing dataset that are indeed hoaxes. 

\end{itemize}
The hoax and nonhoax recalls, together, give a picture of the classification correctness for each of the two classes.  
Generally, we see that the best-performing algorithm is the simple logistic regression based on individual users and words, trained on the whole dataset. 
We note that the topic modeling algorithm is not competitive with the one based on simple logistic regression. 
What is more, the benefit of the algorithm based on simple features over topic modeling is maintained in spite of the fact that the topics are generated over the union of the testing and training sets, as explained in Section~\ref{sec-training-topic-modeling}.
We also see that using textual information helps mostly for URLs that are shared only once.
Text information alone yields inferior results, chiefly due to the low precision of the resulting classifiers.

However, these overall results tell only half of the story. 
For the true insight, we need to look at the outcome of the algorithms on individual news sites.

\subsection{A look at individual news sites}

To gain a better understanding of the performance of the algorithms, it is useful to look at the classification results for individual news sites, reported in Table~\ref{table-news-sites}. 
The results indicate that if we train on the full dataset, the percentages of major news sites URLs that the algorithms label as hoaxes are rather high. 
For instance, about 2\% of URL of the Washington Post are classified as hoaxes, and so are 1\% of the URLs from Reuters and The New York Times.  
In our preliminary experiments, users were negatively impressed when the system labeled as possible hoaxes news that were obviously reliable and came from news sites with strong editorial controls. 
Precision of hoax detection on mainstream news sites was highly valued by our users, much more so than missing the occasional hoax that was shared only a few times (and was thus, likely, a fairly obscure news piece).
Compounding the problem, these are among the largest and most shared news sites, so that even a small percentage of errors translates into a large number of URLs labeled as hoaxes.
For this reason, we find the results obtained with the Min-2 training set to be preferable in practice.

\begin{table*}
\begin{center}
\tiny\begin{tabular}{l|r|r||r|r||r|r||r|r||r|r||r|r||}
 & \multicolumn{4}{c||}{LR-UT} & \multicolumn{4}{c||}{LR-U} & \multicolumn{4}{c||}{Harmonic} \\
 & \multicolumn{2}{c||}{Full Train} & \multicolumn{2}{c||}{Min-2 Train} & \multicolumn{2}{c||}{Full Train} & \multicolumn{2}{c||}{Min-2 Train} & \multicolumn{2}{c||}{Entire, 1x} & \multicolumn{2}{c||}{Entire, 4x}\\
 & OS & MC & OS & MC & OS & MC & OS & MC & OS & MC & OS & MC \\ \hline
nytimes.com                 & 0.88    &   0.83    &   0.42    &   0.35    &   0.83 &  0.73   & 0.19    &   0.14  & 4.10  &  3.95 & 1.37 & 1.18 \\
theguardian.com             &   1.16    &   1.02    &   0.48    &   0.40    &   0.97 &  0.93   & 0.29    &   0.16  & 5.13  &  4.81 & 1.75 & 1.60 \\
huffingtonpost.com          &   2.61    &   1.96    &   1.12    &   0.65    &   1.43 &  1.40   & 0.53    &   0.46  & 7.91  &  7.24 & 2.72 & 2.36 \\
washingtonpost.com          &   1.92    &   2.12    &   0.77    &   0.80    &   3.07 &  2.98   & 0.38    &   0.33  & 7.32  &  7.46 & 2.45 & 2.10 \\
arxiv.org                   &   0.08    &   0.08    &   0.00    &   0.00    &   0.11 &  0.11   & 0.05    &   0.06  & 0.59  &  0.45 & 0.19 & 0.22 \\
usatoday.com                &   1.09    &   1.13    &   0.56    &   0.42    &   0.84 &  0.63   & 0.17    &   0.14  & 4.43  &  4.23 & 1.48 & 1.12 \\
foxnews.com                 &   1.91    &   2.01    &   1.08    &   0.91    &   3.07 &  2.77   & 1.42    &   0.78  & 40.43 &  37.34 & 8.05 & 5.67 \\
nypost.com                  &   3.42    &   3.20    &   1.99    &   1.71    &   3.70 &  3.36   & 1.12    &   1.28  & 17.94 &  17.01 & 4.90 & 3.92 \\
thehill.com                 &   3.39    &   3.43    &   1.79    &   1.60    &   4.64 &  3.55   & 1.91    &   1.09  & 14.71 &  14.09 & 3.96 & 3.31 \\
latimes.com                 &   0.21    &   0.17    &   0.13    &   0.13    &   0.38 &  0.42   & 0.17    &   0.17  & 2.71  &  2.75 & 0.83 & 0.56 \\
breitbart.com               &   5.10    &   3.39    &   3.73    &   1.79    &  11.04 &  9.37   & 6.32    &   4.76  & 77.96 &  82.20 & 20.75 & 13.05 \\
cbsnews.com                 &   1.53    &   1.88    &   0.57    &   0.44    &   1.62 &  1.22   & 0.52    &   0.39  & 11.07 &  10.46 & 3.92 & 3.36 \\
reuters.com                 &   1.16    &   1.11    &   0.44    &   0.44    &   1.07 &  0.93   & 0.62    &   0.36  & 3.54  &  3.26 & 1.29 & 0.98 \\
dailycaller.com             &   25.00   &   41.38   &   16.76   &   32.07   &  50.58 &  56.59  & 39.63   &   50.58 & 86.41 &  85.99 & 29.07 & 20.77 \\
townhall.com                &   31.41   &   16.74   &   21.33   &   14.13   &  38.88 &  23.49  & 28.71   &   16.2  & 78.53 &  76.85 & 31.31 & 18.96 \\
truthfeednews.com           &   21.31   &   2.95    &   16.8    &   2.33    &  18.82 &   1.87  & 15.71   &   2.64  & 98.28 &  97.58 & 81.92 & 35.75 \\
hotair.com                  &   9.40    &   2.03    &   2.99    &   0.96    &  15.60 &  10.15  & 10.58   &   7.48  & 87.38 &  86.22 & 11.40 & 7.01 \\
freedomdaily.com            &   88.89   &   87.96   &   89.81   &   86.11   &  81.48 &  78.70  & 79.63   &   77.78 & 95.88 &  96.30 & 79.42 & 71.60 \\
conservativedailypost.com   &   93.20   &   92.37   &   95.88   &   95.46   &  84.54 &  81.03  & 79.38   &   73.61 & 98.18 &  98.18 & 92.66 & 87.89 \\
lucianne.com                &   15.65   &   96.56   &   2.67    &   96.56   &  89.31 &  96.56  & 77.48   &   96.56 & 99.14 &  98.45 & 26.42 & 34.37 \\
redstate.com                &   39.23   &   14.67   &   38.92   &   9.57    &  39.23 &  18.18  & 26.48   &   11.96 & 78.81 &  71.72 & 23.60 & 10.77 \\
theblaze.com                &   43.06   &   24.07   &   37.96   &   14.81   &  16.67 &  12.50  & 5.09    &   4.63  & 76.01 &  73.99 & 20.61 & 11.92 \\
newsbusters.org             &   18.22   &   21.78   &   14.22   &   21.78   &  29.33 &  31.56  & 20.89   &   31.11 & 89.98 &  89.74 & 17.53 & 11.20 \\
zerohedge.com               &   28.18   &   19.86   &   18.24   &   6.47    &  31.18 &  23.33  & 16.40   &   9.24  & 80.24 &  67.62 & 45.88 & 17.12 \\
\end{tabular}
\end{center}
\caption{Percentage of URLs that are classified as hoaxes for some news sites, including the top news websites of Table~\ref{table-dataset-composition}.
Min-2 Train is the training set consisting of URLs that were shared at least twice; Full Train is the full training set.
OS stands for Opensources ground truth; MC stands for Metacert ground truth.
}
\label{table-news-sites}
\end{table*}

Further, we note that not all sites that have a high percentage of URLs labeled as hoaxes in Table~\ref{table-news-sites} are known for their high standards of journalistic accuracy and editorial controls. 
Indeed, the reputation systems seem to be able to use the Opensources and Metacert list to discover other unreliable sites.
This occurs because the reputation systems assign low reputation (low factors) to the users who share known unreliable news, and then follow those users to other news they have shared. 
As a consequence, the precision figures reported in Table~\ref{table-res-summary} may be underestimates: many of the URLs labeled as hoaxes that belong to sites beyond those in the two ground-truth lists may indeed be hoaxes or misleading news. 

\subsection{Cross-ground-truth prediction}

\begin{table}
\begin{center}
\begin{tabular}{l|l|l|c||r|r|r}
   &       &       & N. of   & \multicolumn{3}{c}{Detection} \\
   &       &       & \multicolumn{1}{c||}{URLs in} & \multicolumn{1}{c|}{Direct} &    & \multicolumn{1}{c}{Suspicious} \\
GT & Mtd.  & Train & diff    & \multicolumn{1}{c|}{URL}  & \multicolumn{1}{c|}{Site} & \multicolumn{1}{c}{URL} \\ \hline \hline
OS & LR-UT & Min-2 & 2530 & 13.9 & 75.0 & 93.2 \\
OS & LR-U  & Min-2 & 2530 & 32.7 & 66.7 & 89.1 \\
MC & LR-UT & Min-2 & 4178 &  9.5 & 56.7 & 70.0 \\
MC & LR-U  & Min-2 & 4178 &  9.1 & 56.7 & 68.2 \\ \hline

OS & LR-UT & Full & 2530 & 21.7 & 91.7 & 95.5 \\
OS & LR-U  & Full & 2530 & 42.6 & 75.0 & 93.3 \\
MC & LR-UT & Full & 4178 & 15.0 & 76.7 & 88.9 \\
MC & LR-U  & Full & 4178 & 18.3 & 76.7 & 83.9 \\ 
\end{tabular}
\end{center}
\caption{Cross dataset detection percentages of fake news and misleading URLs, starting from the ground truth (GT) Opensources (OS) and Metacert (MT).}
\label{table-cross-datasets}
\end{table}
% Luca_good-trainonall-paper-URLs

To validate the ability of the reputation systems for fake news to discover additional unreliable news and sites, we conducted an experiment.
The two ground truths at our disposal, which have been developed independently, share only 331 of 581 (Opensources) and 500 (Metacert) sites respectively.  
Thus, we can ask the question: starting from one ground truth, how good are our reputation systems at discovering the additional URLs and sites in the other ground truth? 
From Table~\ref{table-news-sites}, we see that sites with over 5\% hoax detection are of uncertain quality, worthy of manual investigation by a human. 
Taking 5\% as the threshold for fake or misleading, and starting from a ground truth, we measured:
\begin{itemize}
\item {\em Direct URL detection:} what percentage of URLs appearing only in the other ground truth do we label as hoaxes? 
\item {\em Site detection:} how many sites of the other ground truth with at least 20 URLs in the dataset do we detect as fake or misleading? 
\item {\em Suspicious URL detection:} what percentage of URLs appearing only in the other ground truth belong to suspicious sites? 
\end{itemize}
For site detection, the limitation of sites with at least 20 URLs removes both inactive sites, and sites on which we have so few URLs that computing statistics is unreliable.
The results are given in Table~\ref{table-cross-datasets}. 
As we see, direct URL detection is low: this is in line with the results in Table~\ref{table-news-sites}, where we see that only part of URLs belonging to other fake or misleading sites are flagged as hoaxes.
On the other hand, the detection of fake or misleading URLs is high, above 80\% using full training sets, and above about 70\% for the more conservative Min-2 training set. 
We observe that the Opensources ground truth is more predictive of the Metacert ground truth than the other way round.

\section{Discussion}

We believe that the challenge of building a reputation system for online news sources consists in creating an algorithm that is able to flag a large portion of fake news, while only minimally affecting mainstream news coming from publications with high journalistic standards and editorial controls. 

We obtained the best results via logistic-regression approaches trained on URLs shared via Twitter that were shared more than once, considering users and title/description words as features. 
These models had false positive rates for mainstream news sites and they were generally below 1\%, as can be seen from Table~\ref{table-news-sites}, while yielding an 85\% recall for low-quality news URLs shared more than 10 times.

Algorithms including only users as features worked almost as well as algorithms that included also title and description as features, except on URLs that were shared only once. 
% We believe that this class of algorithms can be used as the foundation of practical news reputation systems. 
Our experiments also indicate that algorithms that treat individual users as topics, such as the logistic and crowdsourcing ones, offer superior performance compared to those that consider common features of users as topics.
We believe this likely happens because on Twitter some users acts as indicators for particular types of news, systematically (re)tweeting a large portion of them. 
Indeed, we have reasons to suspect that some of the users having large influence (weights) in the logistic-regression based methods may be bots and this is something we intend to explore in future work.
We note that our results on topic modeling are preliminary: there may well be more sophisticated ways of applying topic modeling that lead to superior results.
% warn the reader that our results should be considered prelimnary since, we believe, there are more sofisticated ways to implement topic modeling and we plan to explore them in the future.

The ``harmonic'' crowdsourcing-inspired approach worked fairly well.
Its main benefit seems to lie in its ability to spread information from a ground truth to a much wider set of news items and users. Another benefit is that, uniquely among the algorithms we considered, it supports incremental learning. In Table~\ref{table-news-sites}, we note that the harmonic method leads to the least differences in classification starting from our two different ground truths. 
The harmonic method however has two weaknesses.
First, it is based in a very specific graph algorithm that cannot be extended with additional features in the same way as other machine-learning algorithms can.
Second, tuning the harmonic method to deliver good results is a very time-consuming process. 
In most machine-learning algorithms, one can start with small training and testing sets, experiment with hyper-parameters such as class weights, and then run on larger datasets. 
The harmonic algorithm, on the other hand, relies heavily on the connectivity structure of the graph, and it is not obvious how to properly sub-sample the graph to conduct small-scale experiments whose results can be extended to the complete graph.

Our results also highlight an interesting ability of the reputation algorithms considered, namely, their ability to identify fake and misleading news sites that were not known to us in advance.
Indeed, some of the news sites in Table~\ref{table-news-sites} first came to the attention of the authors when looking for false positives.
This suggests an additional use of reputation systemsm for news: as a tool that helps human moderators and fact checkers find more sites that deserve inspection.

\begin{comment}
\subsection*{\tt truthvalue.org}

As a real world application, to help consumers navigate the modern world of fake news, we have exposed some of our algorithms as a service on the web site \url{https://truthvalue.org/}. This is a real-time system that continusly collects twitted URLs and scores them. It produces a dynamic list of recent trending news that have a high reputation and/or a low reputation. The system also allows user to query our database and express their vote about individual news URLs. The users's vote is then weighted according to the reputation they earned (positive or negative).
\end{comment}

\subsection*{Acknowledgements.}
We thank Paul Walsh, CEO of Metacert, for making available to us the Metacert dataset of low-quality sites.

\bibliographystyle{ACM-Reference-Format}
\bibliography{hoax}

\end{document}